\pdfoutput=1
\documentclass[11pt]{article}
\usepackage[a4paper,margin=1in]{geometry}
\usepackage[T1]{fontenc}
\usepackage[utf8]{inputenc}
\IfFileExists{lmodern.sty}{\usepackage{lmodern}}{}
\usepackage{microtype}
\usepackage{booktabs}
\usepackage{tabularx}
\usepackage{array}
\usepackage{graphicx}
\usepackage{amsmath}
\usepackage{amssymb}
\usepackage{threeparttable}
\usepackage{enumitem}
\usepackage{natbib}
\usepackage[hidelinks]{hyperref}
\hypersetup{
    pdftitle={Beyond AI Literacy: A Structured Review and Exploratory Meta-Analysis of Measures for Competent Generative-AI Use},
    pdfauthor={Daniele Veri}
}
\usepackage{xcolor}
\usepackage{caption}
\usepackage{longtable}
\usepackage{pdflscape}
\usepackage{placeins}

\newcolumntype{Y}{>{\raggedright\arraybackslash}X}

\title{Beyond AI Literacy: A Structured Review and Exploratory Meta-Analysis of Measures for Competent Generative-AI Use}
\author{Daniele Ver\'i\\\texttt{daniele.veri.pe@gmail.com}}
\date{9 September 2026}

\begin{document}
\maketitle

\begin{abstract}
AI-literacy measures now include self-report questionnaires, objective tests, and instruments for critical oversight and reliance. Their different targets complicate assessment of competent generative-AI use in work settings. We conducted a structured, seeded evidence review anchored in the 2024 COSMIN-based review of AI-literacy scales, with a targeted update through 17 August 2026. The synthesis covers 24 focal empirical publications plus the prior review and organizes reported measurement content into four domains: knowledge and use, epistemic oversight, reliance calibration, and operational control of tool-using agents. An exploratory meta-analysis of three directly reported subjective-objective correlations, all from one research program, produced a REML pooled $r=.055$ with a Hartung-Knapp 95\% confidence interval of [-.047, .156]. The model uses a combined reported $N=2{,}765$; the largest study has an unresolved discrepancy between its reported correlation and $p$-value, so its weighting requires caution. Adding a synthetic mean of 12 cross-factor correlations from a fourth study yielded $r=.079$, 95\% CI [-.025, .181]. This composite sensitivity addresses a broader comparison than the direct-effect analysis. The small, concentrated evidence base does not establish a population correlation or validate workplace cutoffs, and provides no basis for treating self-ratings as interchangeable with performance scores. Objective instruments such as AICOS-S and GLAT assess foundation knowledge, while other instruments address verification, reliance, trust, and dependency. No validated individual-level instrument in the focal corpus tests the full combination of agent scope, permissions, recovery, state isolation, independent review, and evidence-based closure considered here; some measures cover subsets. We propose a four-layer workplace battery and non-compensatory decision rules as designs for validation. The reviewed evidence motivates separate assessment targets but does not establish the superiority of this battery or its gates.

\end{abstract}

\section{Introduction}

Competent AI use now includes decisions that earlier AI-literacy frameworks did not need to test. Those frameworks asked whether a person could recognize AI, understand broad mechanisms, use applications, evaluate outputs, and reason about ethics \citep{long2020,ng2021}. Those domains remain useful, but tool-using assistants add decisions about access, state changes, execution, and evidence.

A person can answer conceptual questions about large language models and still make poor operational decisions. Examples include accepting an unsupported claim because the prose sounds convincing, granting a tool more access than a task requires, allowing execution to continue after the plan has changed, repairing a corrupted session instead of returning to a known-good state, or treating an agent's completion report as proof that required checks ran. These behaviors sit between AI literacy, critical thinking, reliance, human factors, and professional task competence. Measurement research has addressed each neighboring area in part, but the boundaries between them remain loose.

A persistent methodological split separates self-report from performance measures. Self-report instruments ask people to judge their own capability; performance tests ask them to answer questions or complete tasks with scorable outcomes. In a systematic review completed in mid-2024, \citet{lintner2024} identified 16 AI-literacy scales represented by 22 validation or revalidation studies. Thirteen scales were self-report instruments and three were performance-based. Structural validity and internal consistency received the strongest support across the set, while content validity, reliability, construct validity, responsiveness, interpretability, and feasibility were tested less consistently. Cross-cultural validity and measurement error had not been examined for the included scales.

Work published in 2025 and 2026 broadened that measurement base. GLAT introduced a 20-item performance test focused on generative AI \citep{jin2025}. AICOS and its 12-item short form extended objective assessment to a heterogeneous adult population and included a Generative AI dimension \citep{markus2025,markus2026}. A performance-based adult knowledge scale introduced an explicit epistemic-knowledge dimension \citep{klein2026}. Other teams developed measures for critical thinking during GenAI use \citep{lau2026}, reliance during problem solving \citep{hou2025}, AI information practices \citep{alon2026}, trust \citep{mcgrath2025}, and dependency \citep{goh2025}. These instruments create a richer measurement toolkit, but they also make construct selection harder.

This paper asks four questions:

\begin{enumerate}[label=\textbf{RQ\arabic*.},leftmargin=*]
    \item Which validated instruments now cover the knowledge, evaluation, and use competencies usually grouped under AI or GenAI literacy?
    \item How well do subjective assessments of AI literacy align with objective performance measures when both are collected in the same sample?
    \item Which standardized instruments measure epistemic oversight and reliance, rather than literacy alone?
    \item Which competencies required to supervise tool-using AI agents remain outside validated individual-level measurement?
\end{enumerate}

We treat the registered systematic review by \citet{lintner2024} as the historical baseline and update the evidence relevant to the four questions above. The quantitative synthesis is deliberately narrow: it pools only same-sample associations between perceived and objectively demonstrated AI literacy. Other evidence is synthesized by construct because the instruments differ in item format, population, latent structure, and validation target, making pooled reliability or proficiency estimates hard to interpret.

\section{Conceptual frame}

\subsection{Four measurement domains}

\begin{figure}[htbp]
    \centering
    \includegraphics[width=0.96\textwidth]{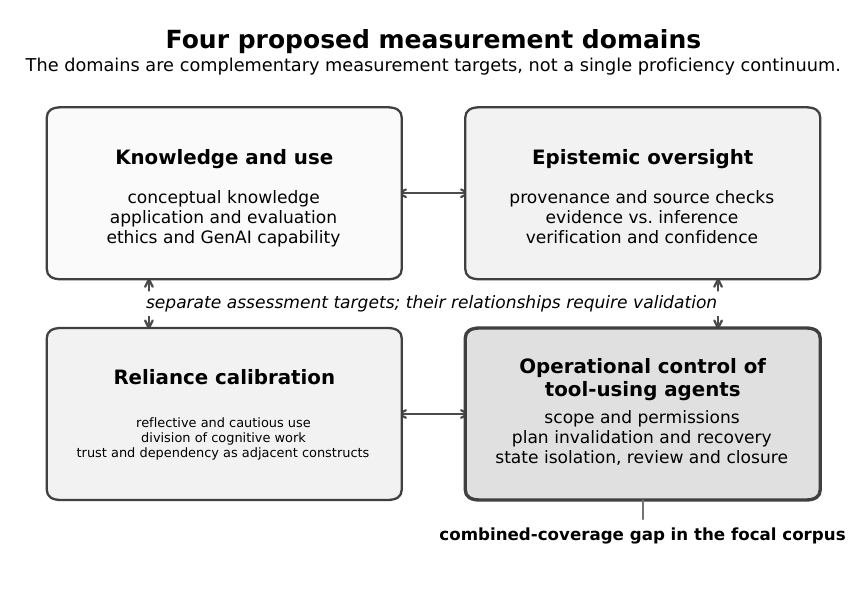}
    \caption{Four-domain framework used to organize the reviewed measures. The domains are treated as complementary measurement targets rather than interchangeable levels of one proficiency continuum. Operational control of tool-using agents is highlighted because no focal instrument directly covered the full set of scope, permission, recovery, state, review, and closure decisions.}
    \label{fig:domains}
\end{figure}

We organize the reviewed instruments into four domains, ordered by their proximity to operational work (Figure~\ref{fig:domains}).

\paragraph{Knowledge and use.}
This domain covers recognition of AI, conceptual and procedural knowledge, application, evaluation, ethics, and GenAI-specific capabilities. It includes both perceived capability measures, such as AILS, SNAIL, and MAILS \citep{wang2023,laupichler2023,carolus2023}, and performance tests such as AILIT, GLAT, AICOS, AICOS-S, SAIL4ALL, the Chiu et al. test, and the scale by Klein-Avraham et al. \citep{hornberger2023,jin2025,markus2025,markus2026,soto2025,chiu2024,klein2026}.

\paragraph{Epistemic oversight.}
This domain concerns what users do when an AI output makes a claim: checking provenance, testing factual support, separating evidence from inference, noticing missing support, and deciding how much confidence a result deserves. The Critical Thinking in AI Use Scale directly measures verification, epistemic motivation, and reflection \citep{lau2026}. AILIS covers information assessment, critique, seeking, retrieval, and source identification \citep{alon2026}. Klein-Avraham et al. place epistemic knowledge inside an objective knowledge framework \citep{klein2026}.

\paragraph{Reliance calibration.}
This domain concerns how people distribute cognitive work between themselves and an AI system. Hou et al. identify reflective, cautious, thoughtless, and collaborative reliance behaviors during problem solving \citep{hou2025}. Trust instruments such as TIAS and S-TIAS measure a related judgment about system trustworthiness and predict willingness to rely, but trust is not itself a competence score \citep{mcgrath2025}. The Generative AI Dependency Scale describes a maladaptive outcome pattern rather than skilled reliance \citep{goh2025}.

\paragraph{Operational control of tool-using agents.}
Once an AI system can act, a new set of observable decisions appears. A user may need to bound scope and permissions, review a plan before state changes, recognize when new evidence invalidates the plan, choose between continuing, narrowing, rewinding, restarting, or escalating, keep independent review separate from authoring, isolate parallel state, and close work with execution evidence. Human-agent teaming research offers behavior-level metrics for preparation, execution, evaluation, adjustment, and team chemistry \citep{dorneich2023}. A task-oriented workplace assessment has also shown that realistic scenario performance can reveal applied AI literacy missed by generic tests \citep{bogart2025}. Neither source provides a validated individual placement scale for the operational behaviors listed above.

Two recent higher-education frameworks partition the same territory differently and are worth mapping explicitly. The AI Literacy Heptagon distinguishes seven dimensions, including Integration Skills and Legal and Regulatory Knowledge \citep{hackl2026}; the AI and Data Acumen framework crosses seven knowledge dimensions with four proficiency levels \citep{kennedy2025}. Our four domains are cut by measurement claim rather than by curricular content, so Integration Skills is distributed across knowledge and use and operational control depending on whether the target is understanding or execution, and Legal and Regulatory Knowledge falls under knowledge and use unless it is assessed through a decision about permitted action, in which case it belongs to operational control. These frameworks organize curricular content and measurement claims in different ways; their dimensions need not conflict.

These domains should not be collapsed into one continuum without evidence. A person may know a great deal about AI and still rely on it poorly. Another person may show disciplined verification while lacking detailed technical knowledge. A third may manage a tool-using agent safely through a familiar workflow without being able to explain model architecture. Training placement needs these distinctions.

\section{Methods}

\subsection{Review design}

We used a structured, seeded evidence review. The search began with the studies and instruments identified by the registered PRISMA/COSMIN review of AI-literacy scales by \citet{lintner2024}. That review searched Scopus and arXiv through 18 June 2024 and provides the most defensible baseline for the pre-2025 literature. We then conducted a targeted update for work published or available through 17 August 2026.

The update searched publisher records and arXiv for combinations of the following concepts: \textit{AI literacy}, \textit{generative AI literacy}, \textit{objective assessment}, \textit{performance-based assessment}, \textit{self-report}, \textit{critical thinking}, \textit{verification}, \textit{reliance}, \textit{trust}, \textit{dependency}, \textit{information literacy}, \textit{workplace task assessment}, and \textit{human-agent teaming}. Forward and backward links from the focal papers were inspected when they pointed to measurement development or direct modality comparisons. Publisher versions were preferred over preprints when both were available.

\begin{figure}[htbp]
    \centering
    \includegraphics[width=0.96\textwidth]{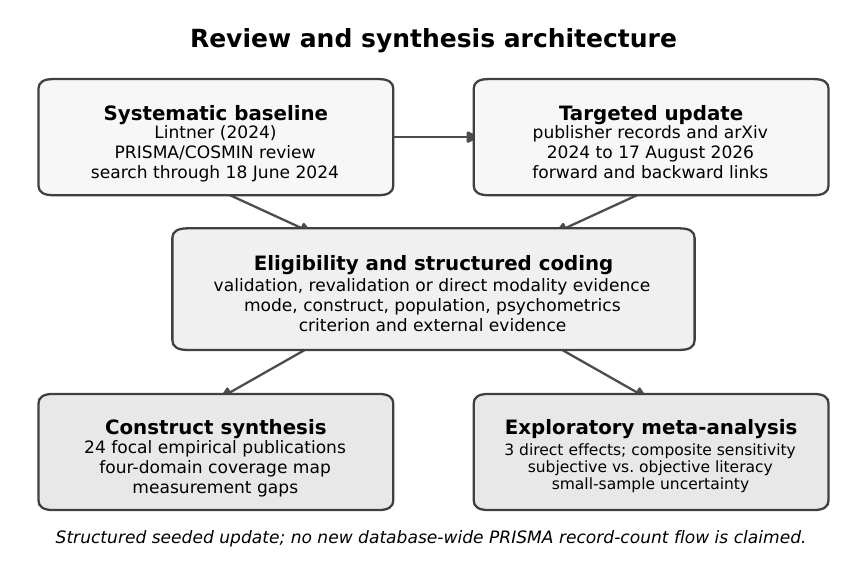}
    \caption{Review and synthesis architecture. The 2024 PRISMA/COSMIN review supplies the systematic baseline; the present study adds a targeted update and structured coding. The synthesis separates a descriptive construct map of 24 focal empirical publications from a primary analysis of three directly reported subjective-objective correlations and a composite-effect sensitivity analysis.}
    \label{fig:reviewarch}
\end{figure}

This update was designed to answer the measurement questions above, not to reproduce a de novo database-wide PRISMA search. We therefore make no claim about a new record-count flow diagram or complete capture of every AI-literacy scale published after June 2024. The search terms and selection rules reported here describe the scope of the update; they do not constitute a reproducible database-wide search protocol. Figure~\ref{fig:reviewarch} summarizes the review and synthesis architecture.

\subsection{Eligibility and coding}

A publication entered the focal synthesis when it met at least one of these conditions:

\begin{enumerate}[leftmargin=*]
    \item it developed, shortened, revalidated, or normed an AI- or GenAI-literacy measure;
    \item it developed a validated measure for critical oversight, reliance, trust, dependency, or AI information practices that bears directly on competent GenAI use;
    \item it compared subjective and objective AI-literacy measures in the same sample; or
    \item it provided a measurement approach for situated workplace AI use or human-agent operational behavior.
\end{enumerate}

For each paper we coded year, instrument, construct, measurement mode, item count, population, sample size, main psychometric evidence, criterion or external evidence, and relevance to the four-domain framework. The focal synthesis contains 24 empirical publications: 22 instrument or assessment reports grouped in Table~\ref{tab:coverage}, the paired-modality study by \citet{zhang2026}, and the oversight study by \citet{dhanorkar2026}. We used the earlier systematic review as the historical anchor. We summarize reported psychometric evidence in the targeted update; we did not conduct a new COSMIN appraisal of these publications.

We treated self-efficacy, trust, reliance, and dependency as adjacent constructs rather than interchangeable labels for competence. T-GASE, for example, states that it assesses perceived capability for text-based GenAI use \citep{durak2026}. TIAS and S-TIAS quantify trust and predict intention to rely \citep{mcgrath2025}. These variables describe beliefs and judgments; neither supplies direct evidence that a person can complete a task correctly.

The coverage map represents the author's interpretive classification of reported instrument content, factors, and assessment tasks. For this descriptive map, \textit{Strong} means that a named dimension or substantial part of the instrument targets the domain; \textit{Partial} means that the instrument addresses a subset of the domain or an explicit adjacent construct; \textit{Weak} means that only a broad use or appraisal facet, or an isolated item, bears on the domain; and \textit{None} means that we identified no operationalized content for it. These are interpretive categories, not a validated coding scale. We report no independent duplicate coding or inter-rater agreement. Coverage describes the target of measurement, not the quality of its validation, the adequacy of a cutoff, or whether the response format demonstrates behavior.

\subsection{Quantitative synthesis}

We restricted the primary quantitative synthesis to directly reported same-sample correlations between subjective AI-literacy ratings and objective AI-literacy performance. \citet{koch2024} reported $r=.21$ between the MAILS total AI-literacy score and ability estimates from the Hornberger AI-literacy test in $N=120$ German-speaking adults. \citet{markus2025} reported $r=.04$ between the 51-item AICOS objective score and MAILS subjective AI literacy in $N=514$. \citet{markus2026} reported $r=.05$ between AICOS-S and the short MAILS, with an overall sample of $N=2{,}131$. All three studies come from one research program and use the MAILS family as the subjective comparator. Shared research origin limits the range of evidence; it does not by itself establish participant overlap.

For AICOS-S, the article reports $r=.05$ and $p=.11$ in Section 3.5, alongside the overall $N=2{,}131$ in Section 2.3 \citep{markus2026}. An unadjusted two-sided Pearson test with that $N$ and $r$ gives $p\approx.021$. The article does not specify a correlation-specific sample size or a correction that resolves this discrepancy. We retained the reported $r$ and used the overall $N$ for the variance, without inferring a replacement $N$ or $r$ from the $p$-value. The resulting weight and combined sample size are conditional on that assumption. We report exclusion of AICOS-S as a sensitivity analysis.

Each direct correlation was transformed as
\begin{equation}
    z_i = \operatorname{atanh}(r_i), \qquad v_i = \frac{1}{N_i-3}.
\end{equation}
We fitted a random-effects model in Fisher-$z$ space using restricted maximum likelihood (REML) to estimate between-study variance. With $w_i=1/(v_i+\hat\tau^2)$ and $\hat\mu=\sum_i w_i z_i/\sum_i w_i$, we used the unmodified Hartung-Knapp interval as the primary interval \citep{knapp2003,rover2015}:
\begin{equation}
    q=\frac{\sum_i w_i(z_i-\hat\mu)^2}{k-1},\qquad
    \hat\mu\;\pm\;t_{k-1,.975}\sqrt{\frac{q}{\sum_i w_i}}.
\end{equation}
We also report the normal-theory interval, $\hat\mu\pm1.96/\sqrt{\sum_i w_i}$. We transformed interval limits back to the correlation scale with $\tanh$. We report between-study variance under both REML and Paule-Mandel, and refit the primary pool using Paule-Mandel as a sensitivity analysis \citep{paule1982}. This matters because the estimators can disagree when the evidence base is small. The model-based prediction interval uses $\hat\mu\pm t_{k-1,.975}\sqrt{\hat\tau^2+1/\sum_i w_i}$. At this $k$, and especially when REML reaches zero, it provides little information about transfer to a new population. We did not test funnel-plot asymmetry or publication bias.

We treated the study by \citet{zhang2026} as a separate exploratory extension. In $N=288$ teachers, the authors reported 12 correlations between four self-report factors and three objective factors, ranging from $.07$ to $.24$, but no total-score correlation. We averaged the Fisher-$z$ transformations of the 12 coefficients and transformed the mean back to $r=.146$. This is an average cross-factor association, including comparisons between different content domains. It is not a correlation between total scores and does not estimate the same comparison as the three direct effects.

The 12 coefficients share participants and factors. Because their covariance matrix is unavailable, we approximated the sampling variance of the composite as
\begin{equation}
    v_{\text{comp}} = \frac{1}{N-3}\cdot\frac{1+(m-1)\rho}{m},
\end{equation}
where $m=12$ and $\rho$ is an assumed common correlation among component estimates. We used $\rho=1$ for the four-effect sensitivity, with $.5$ and $.7$ as alternatives. The value 1 gives the largest variance within this equal-variance, common-dependence approximation; it does not make the composite equivalent to a direct total-score correlation.

Sensitivity analyses comprise leave-one-out refits of the direct-effect pool, the Paule-Mandel refit, addition of the Zhang composite under the three $\rho$ assumptions, and leave-one-out refits of that expanded pool. A descriptive aggregation of the expanded pool reduces each research program to one inverse-variance-weighted Fisher-$z$ mean, with variance $1/\sum_i(1/v_i)$ within each program, before refitting the model across programs. This leaves two aggregate effects and is not a cluster-robust correction or evidence of independent replication. Appendix~\ref{app:meta} reproduces the effect inputs, component correlations, and sensitivity results.

\subsection{Limits of effect availability and extraction}

The quantitative dataset consists of the effects extracted for this analysis, rather than all eligible effects in the literature. Reporting limitations, incomplete eligibility checks, and unfinished numeric extraction restrict its size. The targeted search identified additional candidate records, including a professional-population comparison, for which this review did not complete extraction. We do not classify those records as studies without reported coefficients.

GLAT illustrates a different limitation. \citet{jin2025} administered an objective GenAI test and a self-report ChatGPT literacy scale to the same 83 participants, then regressed task performance on both with visualization literacy and baseline performance as controls. The reported standardized regression coefficient for self-reported literacy was negative and non-significant in that model. The reported regression output does not identify the zero-order correlation between the two literacy scores, so we did not convert it into a meta-analytic effect. Recovery of that effect would require a correlation matrix or additional data from the authors. We also kept self-efficacy comparisons outside the primary pool because self-efficacy is an adjacent construct that would require a separate analysis.

\subsection{Why reliability coefficients were not meta-analyzed}

The instrument set includes dichotomously scored knowledge tests, Likert self-reports, multidimensional batteries, short forms, and behavioral or post-task measures. Some studies report Cronbach's alpha, others composite reliability, omega, IRT information, test-retest coefficients, factor-level reliability, or several of these. A pooled alpha across that mixture would answer no coherent measurement question. We therefore report reliability evidence at instrument level and reserve quantitative pooling for effects with a shared interpretation.

\section{Results}

\subsection{Post-2024 instrument development has expanded objective measurement}

The mid-2024 review described a field dominated by self-report: 13 self-report scales and three performance-based scales \citep{lintner2024}. The post-2024 studies in the focal update add several new performance measures, expanding the objective side of the measurement literature.

GLAT is a 20-item multiple-choice GenAI literacy test validated with 355 higher-education students. Its reported internal consistency was $\alpha=.80$ and $\omega=.81$, and a 2PL IRT model showed good structural fit. GLAT scores predicted performance on GenAI-supported tasks better than perceived ChatGPT proficiency \citep{jin2025}. This criterion result matters because it ties the test to work performed with a generative system rather than to knowledge alone.

AICOS broadened objective measurement to adults. The 51-item instrument covers Apply AI, Create AI, Detect AI, Ethics AI, Generative AI, and Understand AI. In a sample of 514 German-speaking adults it achieved $\alpha=.83$ and composite reliability of $.90$. Scores correlated $r=.58$ with another objective knowledge measure, while the correlation with MAILS subjective AI literacy was $r=.04$ \citep{markus2025}.

AICOS-S reduces the objective test to 12 items and supplies population norms from 2,131 German-speaking adults. The total score showed $\alpha=.71$ and composite reliability of $.83$. A unidimensional CFA fit the data well (CFI=.995, TLI=.994, RMSEA=.014, SRMR=.036). A six-factor model fit too, but the six short subscales had weak reliabilities, so the total score is the defensible interpretation. The objective convergent correlation was $r=.59$; MAILS correlated $r=.05$ \citep{markus2026}. The norming framework places the population median at 8 of 12 items and uses raw scores of 4 and 11 as low and high cut points associated with the 15th and 85th percentile thresholds.

Klein-Avraham et al. extend performance assessment toward epistemic knowledge. Their 26-item adult scale crosses three knowledge types, content, procedural, and epistemic, with three domains, technology, user, and society. The scale was validated with 800 internet-using adults in Israel. Epistemic and society-related knowledge were negatively associated with trust in GenAI, suggesting that more informed users did not simply become more trusting \citep{klein2026}.

Other objective tests remain useful in narrower populations. The Hornberger AILIT targets higher education \citep{hornberger2023}; Chiu et al. validated 25 multiple-choice items with 2,390 students in grades 7 through 9 using Rasch analysis \citep{chiu2024}; SAIL4ALL covers four broad adult knowledge themes and offers both true/false and Likert response formats, while its multidimensional structure argues against a single overall total \citep{soto2025}.

\subsection{Self-report remains useful when the construct is subjective}

The rise of objective tests does not make self-report obsolete. It changes what a self-report score can be claimed to measure.

AILS operationalizes awareness, use, evaluation, and ethics through 12 self-report items \citep{wang2023}. SNAIL measures perceived technical understanding, critical appraisal, and practical application \citep{laupichler2023}. MAILS adds self-efficacy and self-management to AI-literacy facets and supports modular use \citep{carolus2023}; a 10-item short form has received further validation \citep{koch2024}. AILST provides a teacher-specific 36-item measure across AI perception, knowledge and skills, application and innovation, and ethics, validated with 604 respondents \citep{ning2025}. T-GASE narrows the target to confidence in using text-based GenAI and contains 23 items across Application, Expectations, Ethics, and Evaluation \citep{durak2026}. Each can support needs analysis, confidence profiling, or evaluation of perceived change.

The interpretation changes when a self-report score is used as a gate for proficiency. A respondent may be well calibrated, overconfident, or underconfident. The next section quantifies the limited alignment seen in studies that measured both modes.

\subsection{Exploratory meta-analysis: perceived versus demonstrated AI literacy}

\begin{figure}[htbp]
    \centering
    \includegraphics[width=\textwidth]{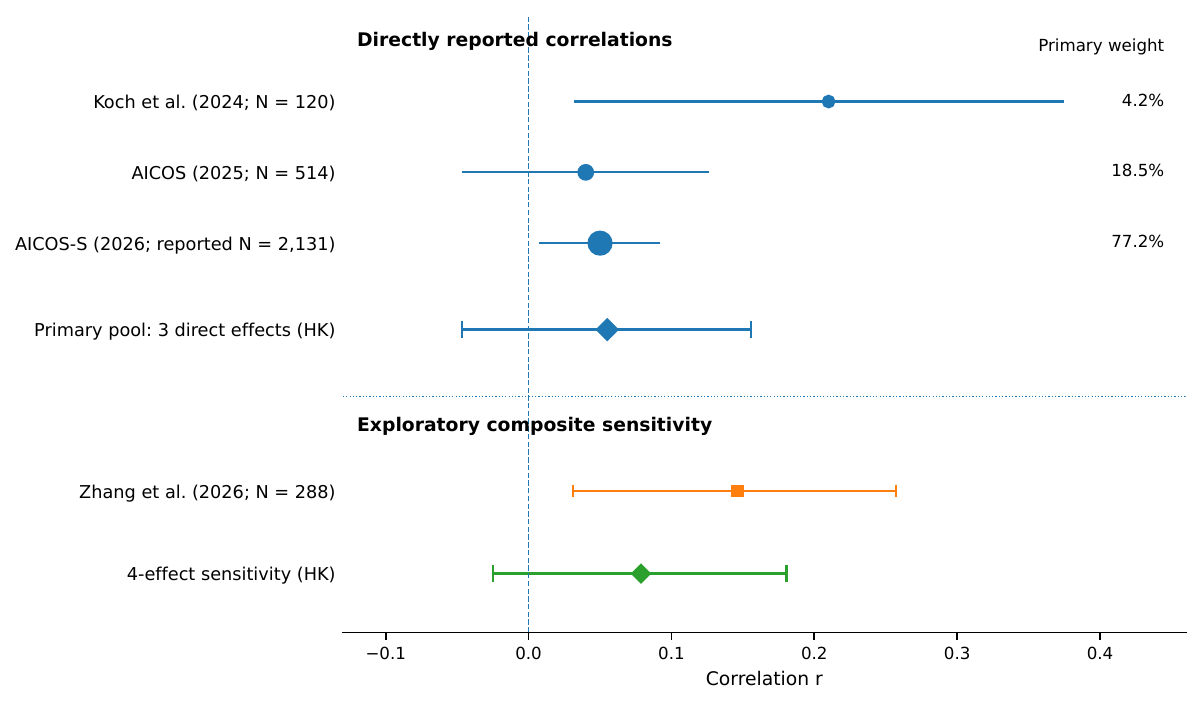}
    \caption{Direct subjective-objective correlations and the separate composite sensitivity. Individual direct-effect markers scale with primary REML weight; the displayed AICOS-S $N$ is the reported overall sample size, with the weighting caveat in Methods. Individual intervals use the Fisher-$z$ normal approximation. Diamonds show pooled Hartung-Knapp 95\% intervals. Zhang's square is the mean of 12 dependent cross-factor correlations, with $\rho=1$ for its approximate variance; it enters only the four-effect sensitivity.}
    \label{fig:forest}
\end{figure}

Figure~\ref{fig:forest} separates the three direct effects from the composite-effect sensitivity. The primary pool uses a combined reported $N=2{,}765$, subject to the AICOS-S sample-size assumption described in Methods. Cochran's $Q=3.094$ on two degrees of freedom yielded $I^2=35.4\%$. REML reached the boundary, $\hat\tau^2\approx0$, while Paule-Mandel gave $\hat\tau^2=.0033$. With three effects, this disagreement cautions against treating the dispersion estimate as stable.

The primary REML pooled correlation was $r=.055$, with a Hartung-Knapp 95\% interval of [-.047, .156] and a normal-theory interval of [.018, .092]. Its model-based prediction interval was [-.027, .136]; the REML boundary estimate makes this interval particularly dependent on model assumptions. Refitting with Paule-Mandel gave $r=.072$ and a Hartung-Knapp interval of [-.113, .252]. The point estimates suggest limited alignment in these samples, while the intervals include zero.

AICOS-S carries 77.2\% of the primary REML weight under the reported overall sample size. Removing it gives $r=.107$, with a Hartung-Knapp interval of [-.747, .828] across the two remaining studies. The other two leave-one-out point estimates are $.048$ and $.106$. These refits describe sensitivity to individual studies; their small-$k$ intervals do not establish a stable population relationship.

Adding the Zhang cross-factor composite yields four effects and a combined reported $N=3{,}053$: $r=.079$, Hartung-Knapp 95\% CI [-.025, .181], normal-theory CI [.021, .136], and model-based prediction interval [-.070, .224]. For this expanded pool, $Q=5.290$, $I^2=43.3\%$, $\hat\tau^2_{\mathrm{REML}}=.0013$, and $\hat\tau^2_{\mathrm{PM}}=.0025$. Varying $\rho$ from 1 to $.5$ changes the point estimate from $.079$ to $.088$. This sensitivity concerns the assumed composite variance; it does not resolve the difference in what the composite measures. In the expanded pool, AICOS-S carries 47.5\% of the weight, and removing it changes $r$ from $.079$ to $.113$. Aggregating the expanded data to two research-program effects gives $r=.084$, with a Hartung-Knapp interval of [-.430, .557]. Appendix~\ref{app:meta} lists the sensitivity results.

The extracted effects give no basis for treating subjective and objective scores as interchangeable in these samples. They do not establish a proficiency cutoff, quantify placement errors, or show that all self-report scales measure the same construct. The workplace proposal below therefore draws on the distinction between measurement targets as well as on this limited association evidence.

The teacher study by \citet{zhang2026} also examined profiles of mismatch. Its 12 cross-factor correlations ranged from $.07$ to $.24$, and latent profile analysis identified patterns of overestimation, underestimation, and alignment. Those profiles provide additional descriptive evidence about self-assessment; they do not turn the cross-factor mean into a total-score correlation.

\subsection{Critical oversight has become measurable}

Objective AI knowledge alone does not capture how a user scrutinizes an answer during use. Two independent instrument-development efforts targeted this distinction within a single year, giving the construct stronger empirical footing than either effort would provide alone.

The Critical Thinking in AI Use Scale measures this distinction directly. Across six studies with $N=1{,}341$, \citet{lau2026} developed a 13-item measure with Verification, Motivation, and Reflection factors. Criterion testing moved beyond correlations among questionnaires: higher scores predicted more frequent and varied verification strategies, greater accuracy in judging the veracity of claims in a naturalistic GPT-powered fact-checking task, and deeper reflection about responsible AI. The criterion task is what distinguishes this instrument, because it ties a self-report score to observed judgements about specific claims.

CAILS approaches a comparable construct from educational technology. Its 24 items span knowledge-related, operational, critical, and ethical dimensions and were validated with 314 first-year student teachers after a 57-participant doctoral pilot. Reliability for the four dimensions ranged from $\alpha=.838$ to $.912$ \citep{ranieri2025}. The critical dimension is a designed component rather than a by-product, so the claim that critical oversight has only recently become measurable would be too strong. What CAILS does not yet have is criterion evidence against observed verification behaviour, which is precisely where Lau et al. add something. For a layered battery, the two instruments provide different evidence: CAILS embeds the construct alongside knowledge and ethics, while the Critical Thinking scale adds a behavioral criterion.

AILIS examines similar behavior from an information-science perspective. Its 39 self-report items cover creation and processing, assessment and critique, seeking and retrieval, and ethics in a sample of 758 adults \citep{alon2026}. Items address missing evidence, factual and logical errors, source identification, currency, attribution, privacy, and choosing when another source or tool is more appropriate. AILIS therefore maps well to information-intensive work, although its score remains a report of one's own practices.

Klein-Avraham et al. provide an objective counterpart by separating epistemic knowledge from content and procedural knowledge \citep{klein2026}. Taken together, these measures distinguish knowledge about how AI-generated information should be interpreted from the disposition or reported practice of verifying it.

\subsection{Reliance, trust, and dependency should remain separate}

Hou et al. validated a scale for reliance behaviors during GenAI-assisted problem solving. Exploratory factor analysis used 800 responses, followed by confirmatory analyses on a holdout sample of 730 responses and 1,173 responses from a second problem-solving activity. The four factors were reflective, cautious, thoughtless, and collaborative use; overall internal consistency was $\alpha=.84$ \citep{hou2025}. Because the scale describes behavior during problem solving, it sits closer to competent collaboration than a generic trust score.

McGrath et al. validated the 12-item Trust in Automation Scale for contemporary AI applications and developed a three-item S-TIAS \citep{mcgrath2025}. Both scales responded to manipulations of system trustworthiness, and S-TIAS predicted intention to rely. These properties make S-TIAS useful when trust calibration is the research target. A high or low trust score alone cannot establish good judgment. Appropriate reliance depends on whether trust changes with evidence about system performance and whether the user's action matches task risk.

The Generative AI Dependency Scale targets a separate construct. Across six studies with 1,333 participants, it showed a three-factor structure covering cognitive preoccupation, negative consequences, and withdrawal, with $\alpha=.92$ to $.93$ and test-retest ICC=.87 \citep{goh2025}. Dependency was associated with lower task performance and critical thinking. It can serve as an outcome or risk indicator and should remain separate from an AI-literacy total.

\subsection{The operational-agent measurement gap}

The reviewed standardized instruments cover increasingly sophisticated parts of GenAI use, yet they stop before several decisions created by tool-using agents. We found no validated individual-level scale in the focal corpus that directly tests all of the following behaviors:

\begin{itemize}[leftmargin=*]
    \item selecting the smallest permission and action surface needed for a task;
    \item recognizing that new evidence invalidates an approved plan and requires re-planning;
    \item preserving a known-good state and choosing among continuation, narrowing, rewind, restart, and escalation;
    \item separating maker and reviewer roles so that the same reasoning trace does not anchor both;
    \item isolating state when several agents work in parallel and planning reconciliation before execution;
    \item distinguishing an agent's completion claim from independent execution evidence; and
    \item closing work with failed, skipped, and unverified checks visible.
\end{itemize}

Dorneich et al. show that preparation, execution, evaluation, adjustment, and team chemistry can be operationalized as observable human-human and human-agent teaming behaviors \citep{dorneich2023}. Their unit of analysis is team behavior in gameplay, not individual workplace proficiency. Bogart et al. move closer to occupational assessment: in a US Navy robotics training context, a scenario task designed to simulate job use was more informative for applied AI literacy than generic tests adopted or developed for the program \citep{bogart2025}. That work is currently a preprint and does not supply a general normed scale.

The list draws on the conceptual framework and reports of oversight work. \citet{dhanorkar2026} interviewed 17 experienced developers and identified four forms of emergent oversight work: a priori control, co-planning, real-time monitoring, and post hoc review. They also document situated review difficulties and practitioner heuristics used to cope with them. Those observations motivate candidate assessment content. They do not validate the full list as an individual competence construct or establish how each behavior should be scored.

Regulatory and technical work provides related context. ISO/IEC FDIS 42105 addresses human oversight of AI systems \citep{iso42105}. \citet{nannini2026} examine whether oversight remains commensurate with risk across multi-step agent action chains. Their question concerns the adequacy of system oversight, whereas this review concerns measurement of individual competence. The related concerns help motivate assessment development; they do not provide independent confirmation that no individual instrument exists or remove the selection limits of this review.

Work on delegation supplies vocabulary that the construct should adopt rather than reinvent. \citet{south2025} frame agent delegation around authentication, authorization, and auditability and show how natural-language permissions can be translated into auditable access-control configurations. A separate delegation framework by \citet{tomasev2026} decomposes delegation into task allocation plus transfer of authority, responsibility, accountability, role and boundary specification, clarity of intent, and mechanisms for establishing trust. These concepts map onto the scope, permission, review, and evidence-based closure behaviors above and keep the proposed construct commensurable with technical work already under way.

The proposed construct concerns performance under operational constraints. We use \textit{agentic operational competence} as a provisional label for the ability to control authority, state, recovery, review, and evidence while delegating work to a tool-using AI system. The label should be treated as a hypothesis for instrument development, not as an established latent variable.

\section{Cross-instrument synthesis}

Table~\ref{tab:coverage} summarizes the author's descriptive coverage judgments using the categories defined in Methods. These judgments distinguish content coverage from measurement mode and validation quality. A strong content rating does not imply a demonstrated skill or a validated placement decision.

\begin{table}[htbp]
\centering
\footnotesize
\begin{threeparttable}
\caption{Construct coverage of focal instruments}
\label{tab:coverage}
\begin{tabularx}{\textwidth}{@{}Y Y *{4}{>{\centering\arraybackslash}p{1.72cm}}@{}}
\toprule
Instrument & Primary mode & Knowledge \& use & Epistemic oversight & Reliance & Agent operations \\
\midrule
AILS & Self-report & Strong & Partial & Weak & None \\
SNAIL & Self-report & Strong & Partial & Weak & None \\
MAILS / short MAILS & Self-report & Strong & Partial & Partial & None \\
AILIT & Performance & Strong & Partial & None & None \\
Chiu et al. AI test & Performance & Strong & Partial & None & None \\
SAIL4ALL & Performance / Likert & Strong & Partial & None & None \\
GLAT & Performance & Strong & Partial & Partial & None \\
AICOS / AICOS-S & Performance & Strong & Partial & None & None \\
Klein-Avraham et al. & Performance & Strong & Strong & Partial & None \\
T-GASE & Self-report & Strong & Partial & Partial & None \\
AILST & Self-report & Strong & Partial & Weak & None \\
FALCON-AI & Self-report & Strong & Strong & Partial & Weak\tnote{e} \\
CT in AI Use & Self-report + task criterion & Partial & Strong & Partial & None \\
CAILS & Self-report & Strong & Strong & Partial & None \\
AILIS & Self-report & Partial & Strong & Partial & None \\
GenAI Reliance & Post-task self-report & Partial & Partial & Strong & None \\
TIAS / S-TIAS & Self-report & None & None & Partial\tnote{a} & None \\
GenAI Dependency & Self-report & None & Partial & Partial\tnote{b} & None \\
Dorneich team metrics & Observed behavior & None & Partial & Partial & Partial\tnote{c} \\
Bogart task assessment & Scenario performance & Strong & Strong & Partial & Partial\tnote{d} \\
\bottomrule
\end{tabularx}
\begin{tablenotes}[flushleft]\footnotesize
\item[] \textit{Note.} The 20 rows group 22 focal instrument and assessment publications; MAILS and AICOS each combine a full-form and a short-form report. Zhang et al. and Dhanorkar et al. supply the other two focal publications. Ratings describe content coverage, not psychometric quality or suitability for certification.
\item[a] Trust predicts intention to rely, but trust is not a direct measure of appropriate reliance.
\item[b] Dependency represents a maladaptive use pattern, not skilled reliance.
\item[c] Measures team behaviors across human-human and human-agent teams rather than individual workplace competence.
\item[d] Work-task alignment is strong, but the paper is a preprint and the instrument is context-specific.
\item[e] A self-report item refers to customizing an agent. This isolated content supports a weak coverage judgment, not evidence of executing or supervising an agent task.
\end{tablenotes}
\end{threeparttable}
\end{table}

Several assignments require qualification. GLAT and the Klein-Avraham instrument receive partial reliance coverage because their evaluation or epistemic content bears on reliance judgments; neither directly observes reliance calibration. AILS, SNAIL, and AILST receive weak reliance coverage because their broad appraisal and use facets only touch this target. MAILS self-management, information practices in AILIS, and the critical or ethical dimensions of other instruments address subsets or adjacent constructs. The table notes distinguish trust and dependency from skilled reliance, and team or workplace assessment from the full individual agent-control target.

No instrument in the focal corpus covers the full combination of four domains defined here. Some instruments cover subsets of operational control, so the final column does not establish the absence of measures for every individual behavior. The map motivates separate assessment targets; it does not establish that combining these instruments improves workplace decisions.

Measurement mode should follow the claim being made. A self-efficacy scale is appropriate when confidence is the target. A performance test is appropriate when demonstrated knowledge is the target. A role-specific scenario is appropriate when the decision concerns execution under constraints. Problems start when one mode is asked to stand in for another.

Advanced use of tool-using AI calls for assessment at the level of decisions and traces. An answer can be scored correct while the process that produced it violated an information boundary, bypassed a check, or changed an unauthorized artifact. Operational assessment therefore needs artifacts such as source locators, permission choices, state transitions, test results, and recovery decisions in addition to a final answer.

\section{Implications for workplace placement and training evaluation}

\subsection{A proposed layered battery}

\begin{figure}[htbp]
    \centering
    \includegraphics[width=0.98\textwidth]{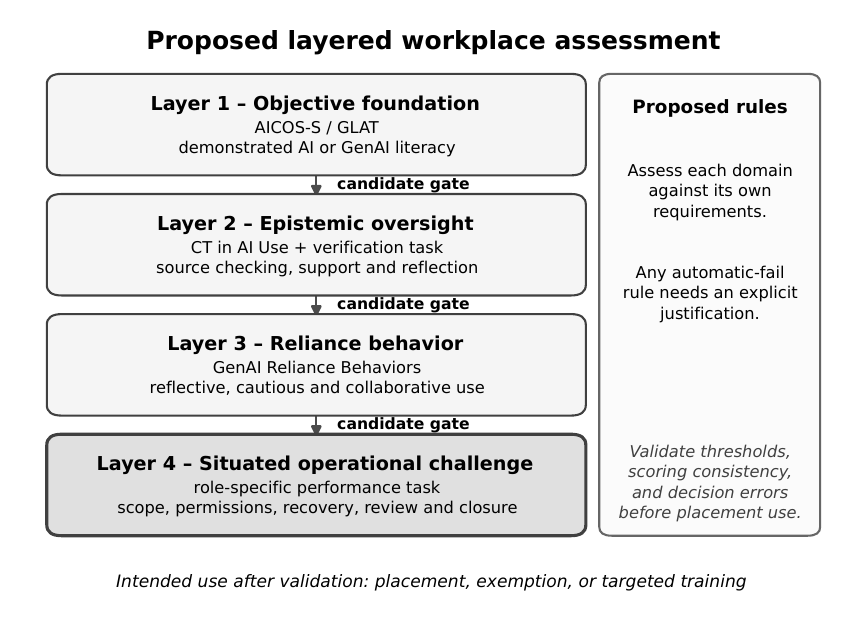}
    \caption{Proposed layered workplace assessment. The layers separate measurement targets; the gates illustrate a non-compensatory decision design for validation. No study in this synthesis tested this full battery, gate thresholds, or its superiority over alternative scoring approaches. The operational challenge requires role-specific development and validation.}
    \label{fig:layered}
\end{figure}

We propose a four-layer workplace assessment that keeps the measurement targets separate (Figure~\ref{fig:layered}). The synthesis motivates this design but does not test its incremental validity against a single score, another battery, or a multidimensional profile. The following layers are candidate components for local validation, not a validated placement protocol.

\paragraph{Layer 1: objective foundation.}
AICOS-S is a candidate for a brief adult baseline: it has 12 objectively scored items, includes GenAI content, and has German-speaking population norms \citep{markus2026}. GLAT is an alternative for a GenAI-focused assessment in populations resembling its higher-education validation samples, with criterion evidence against GenAI-supported task performance \citep{jin2025}. A full AICOS administration covers more items when time permits, although its subscale reliability limits fine-grained diagnosis \citep{markus2025}. These properties do not establish workplace pass thresholds.

\paragraph{Layer 2: epistemic oversight.}
The Critical Thinking in AI Use Scale adds verification, motivation to understand, and reflection, and its criterion task links scores to actual fact-checking behavior \citep{lau2026}. AILIS is useful when source handling and information work dominate the job \citep{alon2026}. Where demonstrated performance matters, a short verification task should accompany either self-report instrument.

\paragraph{Layer 3: reliance behavior.}
The GenAI Reliance Behaviors Scale can describe reflective, cautious, thoughtless, and collaborative patterns after a realistic problem-solving task \citep{hou2025}. S-TIAS can be added when trust calibration is itself a research question, but it should remain separate from proficiency \citep{mcgrath2025}.

\paragraph{Layer 4: situated operational challenge.}
The final layer should reproduce the decisions that matter in the role. A software task might require inspection before modification, bounded permissions, plan review, test evidence, recovery from a faulty assumption, and independent review. A research task might require source authority, provenance, separation of observation from interpretation, and treatment of participant data. QA and operations tasks would use different action surfaces. The task-oriented assessment reported by \citet{bogart2025} motivates role-aligned scenarios. \citet{song2026} report initial validation of a faculty self-report instrument in 269 respondents, crossing three literacies with four work domains. Their preprint offers an example of role-embedded measurement, while Bogart et al. assess scenario performance. Neither establishes a validated general test of the full agent-control behavior set proposed here.

Separate results could help distinguish training needs. A participant with high perceived competence and weak demonstrated knowledge might need foundation work; a participant with strong knowledge and weak verification might need different training. Success on questionnaires or knowledge tests would not establish success on an operational challenge involving authority or state. These are illustrative interpretations. The proposed battery has not yet shown that it improves placement accuracy or training outcomes.

\subsection{Non-compensatory rules as a design option}

A weighted total can allow strength in one domain to offset weakness in another. A non-compensatory rule can instead require evidence for each task-relevant requirement. For example, a knowledge score would not offset granting write access during a read-only task, and a high self-rating would not replace successful verification of a claim. These examples express design priorities; the review does not estimate how often either scoring approach would misclassify participants.

Proficiency-level matrices offer another way to organize assessment, including the dimensions and levels described by \citet{kennedy2025}. A matrix need not be compensatory: assessors can retain separate requirements for each dimension. The choice concerns the aggregation and decision rules, not the use of levels itself. The reviewed studies do not compare matrices, gates, and weighted totals for workplace placement.

A candidate gate design could require a foundation threshold, a verification threshold, and evidence of controlling scope, action, recovery, review, and closure in a situated task. Designers could specify automatic-fail behaviors for safety or integrity failures and score ordinary errors within a rubric. Before using such rules for consequential placement, researchers would need to justify the thresholds, establish scoring consistency, examine false-pass and false-fail decisions, and test whether the rules improve on simpler alternatives.

\subsection{Candidate operational scenarios and validation needs}

Advanced questions should require coordination of several principles. Vocabulary recognition is easy to coach and easy to game. A better scenario changes one premise during execution and asks what the participant does with scope, evidence, and authority. Another presents a known-good checkpoint after a wrong assumption and asks whether to repair, rewind, restart, or escalate. A reviewer scenario can test whether the participant separates fresh evaluation from the original agent's rationale. Parallel-work scenarios can test state isolation and reconciliation.

Such items are situational judgment tasks. Their validity will depend on job analysis, expert review, scoring consistency, difficulty, and predictive evidence. Standard psychometric work remains necessary before a local challenge can be called a standardized instrument. The studies reviewed here motivate situated assessment, but do not validate this item format or a ready-made item bank for the proposed agent-control construct.

\section{Limitations}

This review has five limits.

First, the update is structured and seeded rather than a de novo systematic database review. Lintner's 2024 review supplies the systematic baseline; our search extends it through targeted publisher and arXiv retrieval. Newly published scales outside those paths may be missing.

Second, the focal publications span general adults, university students, K-12 students, teachers, and specialized training contexts. Measurement properties do not transfer automatically across these groups. A norm from German-speaking adults cannot be applied as a workplace cutoff in another country without local evidence. The primary pool covers German-speaking adults; the expanded sensitivity adds teachers. Employed adults and in-service teachers are represented in this evidence, but the studies do not validate the proposed workplace placement decisions. Extracting further professional-population comparisons would broaden the samples without establishing cross-role transfer on its own.

Third, AI-related item content ages quickly. Questions about model capabilities, prompting, or tool behavior can change difficulty as products change. Stable assessment should emphasize durable principles where possible and maintain a versioned item-review process for tool-specific content.

Fourth, all three primary effects come from one research program using the MAILS family. AICOS-S carries 77.2\% of primary REML weight, conditional on using its overall reported sample size despite the unresolved $r$/$p$ discrepancy. The primary interval includes zero, and the heterogeneity estimators disagree. The expanded sensitivity adds a cross-factor composite that measures a different comparison, and its interval also includes zero. No analysis corrects for unreliability, differences in content coverage, or range restriction. These estimates describe a small, selected evidence base and do not validate score interchangeability or placement rules. Incomplete extraction of additional candidates further limits claims about the wider literature.

Fifth, no instrument in the focal corpus tests the full combination of agent-control behaviors proposed here, although some cover subsets. The four-domain map, layered battery, gate rules, and label \textit{agentic operational competence} are organizing proposals. The coverage ratings reflect one author's interpretation, without independent duplicate coding or a new COSMIN appraisal. Researchers would need content validation, scenario development, pilot testing, reliability analysis, criterion evidence, and cross-role comparisons before using the proposed assessment for standardized certification.

\section{Research agenda}

The next measurement step should connect psychometrics to traces of real AI-assisted work.

A first study could develop a situational judgment test from critical incidents collected across software development, QA, operations, research, and information-intensive office work. Subject-matter experts would rate candidate responses for safety, effectiveness, and evidence quality. Think-aloud pilots could determine whether items test the intended decision rather than obscure domain knowledge.

A second study could pair the new scenario test with AICOS-S or GLAT, the Critical Thinking in AI Use Scale, and the GenAI Reliance Behaviors Scale. This design would test convergent and discriminant relations among knowledge, epistemic oversight, reliance, and operational control. If the domains are distinct, a multi-factor model should fit better than a single generic ``AI competence'' factor.

A third study should collect behavioral traces from a sandboxed tool-using agent. Candidate markers include permission requests, plan changes, source locators, checkpoint use, unauthorized-scope attempts, restart or rewind decisions, verification commands, reviewer separation, and completion-report accuracy. These traces could supply criterion evidence that questionnaire-only studies cannot provide.

A fourth study should test transfer. Placement instruments become useful when they predict later behavior: fewer unsupported claims, fewer unauthorized actions, more complete evidence, better recovery, and lower rework on comparable tasks. Thirty- and sixty-day samples of completed work would provide a stronger criterion than immediate post-course satisfaction.

\section{Conclusion}

The reviewed instruments assess different components of competent GenAI use. Objective tests such as AICOS-S and GLAT assess demonstrated foundation knowledge; self-reports assess perceptions, confidence, and reported practice. The exploratory primary analysis of three direct subjective-objective correlations yielded $r=.055$, with a Hartung-Knapp interval that includes zero. All three effects come from one research program, and the largest study's weighting depends on an unresolved reporting discrepancy. Adding a cross-factor composite from another study gives $r=.079$, also with an interval that includes zero. These results provide no basis for substituting self-ratings for performance scores, but do not establish a population correlation or a workplace pass threshold.

Other instruments address verification, reflective oversight, reliance, trust, and dependency. No validated individual-level instrument in the focal corpus tests the full combination of scope changes, permission decisions, recovery from faulty state, state isolation, independent review, and evidence-based closure considered here. This is a corpus-bounded finding about combined coverage, not a claim that no measures address its component behaviors.

We propose a workplace battery that keeps objective literacy, epistemic oversight, reliance behavior, and situated operational performance separate. Non-compensatory rules are one candidate approach to using those results. The synthesis motivates these designs; their added value, thresholds, error rates, and transfer across roles remain empirical questions.

\section*{Data availability}

This review uses published aggregate statistics and collected no participant-level data. Appendix~\ref{app:meta} reproduces the quantitative inputs and sensitivity results, including the Zhang component correlations. Table~\ref{tab:coverage} reports the descriptive coverage judgments. No separate data files, analysis scripts, or supplementary verification records accompany this arXiv submission.

\FloatBarrier

\bibliographystyle{plainnat}
\bibliography{references}

\appendix
\section{Quantitative inputs and sensitivity results}
\label{app:meta}

Table~\ref{tab:inputs} lists the three direct effects used in the primary analysis and the composite used in the expanded sensitivity. Table~\ref{tab:zhang} reproduces its components. All correlations are uncorrected for measurement error. The reported sample totals assume that each study's stated $N$ applies to the extracted coefficient; the AICOS-S caveat below requires particular attention.

\begin{table}[htbp]
\centering
\small
\begin{threeparttable}
\caption{Study-level inputs and analytic roles}
\label{tab:inputs}
\begin{tabularx}{\textwidth}{@{}Y r r Y@{}}
\toprule
Study & $N$ used & $r$ & Definition and role \\
\midrule
Koch et al. (2024), Study III & 120 & .210 & MAILS literacy total vs. Hornberger test ability estimate; primary \\
Markus et al. (2025), AICOS & 514 & .040 & AICOS total vs. subjective MAILS; primary \\
Markus et al. (2026), AICOS-S & 2,131\tnote{a} & .050 & AICOS-S total vs. short MAILS; primary \\
Zhang et al. (2026) & 288 & .146 & Mean Fisher-$z$ of 12 cross-factor correlations; sensitivity only \\
\bottomrule
\end{tabularx}
\begin{tablenotes}[flushleft]\footnotesize
\item[a] Overall study $N$, not a separately confirmed pairwise sample size. The source reports $r=.05$, $p=.11$; an unadjusted Pearson test with $N=2{,}131$ gives $p\approx.021$. We preserve the reported correlation and disclose the unresolved discrepancy rather than imputing a replacement value. Exclusion sensitivities appear in Table~\ref{tab:sensitivity}.
\end{tablenotes}
\end{threeparttable}
\end{table}

\begin{table}[htbp]
\centering
\small
\caption{Zhang et al. (2026): component correlations used in the sensitivity composite}
\label{tab:zhang}
\begin{tabularx}{\textwidth}{@{}Y *{3}{>{\centering\arraybackslash}X}@{}}
\toprule
Self-report factor & Objective conceptual understanding & Objective capability evaluation & Objective practical and ethical use \\
\midrule
Concept & .24 & .14 & .14 \\
Ethics & .23 & .17 & .11 \\
Evaluate & .17 & .10 & .07 \\
Use & .18 & .10 & .10 \\
\bottomrule
\end{tabularx}
\smallskip
\parbox{\textwidth}{\footnotesize\textit{Note.} All 12 coefficients share the same $N=288$. Their Fisher-$z$ mean back-transforms to $r=.146252$. The composite is an average cross-factor association, not a total-score correlation. Its approximate variance is $1/285=.003509$ at $\rho=1$, $.002544$ at $\rho=.7$, and $.001901$ at $\rho=.5$.}
\end{table}

\begin{table}[htbp]
\centering
\small
\begin{threeparttable}
\caption{Primary model and sensitivity analyses}
\label{tab:sensitivity}
\begin{tabularx}{\textwidth}{@{}Y r r r@{}}
\toprule
Analysis & $k$ & Pooled $r$ & Hartung-Knapp 95\% CI \\
\midrule
Primary: three direct effects, REML & 3 & .055 & [-.047, .156] \\
Primary with Paule-Mandel variance & 3 & .072 & [-.113, .252] \\
Direct effects without Koch et al. & 2 & .048 & [-.002, .098] \\
Direct effects without AICOS & 2 & .106 & [-.707, .798] \\
Direct effects without AICOS-S & 2 & .107 & [-.747, .828] \\
\addlinespace
Expanded: add Zhang composite, $\rho=1$ & 4 & .079 & [-.025, .181] \\
Expanded, $\rho=.7$ & 4 & .085 & [-.024, .191] \\
Expanded, $\rho=.5$ & 4 & .088 & [-.022, .196] \\
Expanded without Koch et al. & 3 & .058 & [-.032, .147] \\
Expanded without AICOS & 3 & .109 & [-.088, .298] \\
Expanded without AICOS-S & 3 & .113 & [-.098, .314] \\
Expanded without Zhang & 3 & .055 & [-.047, .156] \\
Expanded, aggregated by research program & 2 & .084 & [-.430, .557] \\
\bottomrule
\end{tabularx}
\begin{tablenotes}[flushleft]\footnotesize
\item[] \textit{Note.} All models use REML except the Paule-Mandel row. Leave-one-out analyses of the expanded pool use $\rho=1$. Primary reported $N=2{,}765$; expanded reported $N=3{,}053$. Normal-theory intervals are [.018, .092] for the primary REML model and [.021, .136] for the expanded model. Their model-based prediction intervals are [-.027, .136] and [-.070, .224]. The primary REML heterogeneity estimate is at zero; these intervals should not be treated as reliable coverage statements for new workplace populations. The program aggregation is descriptive, not a cluster-robust adjustment.
\end{tablenotes}
\end{threeparttable}
\end{table}
\FloatBarrier

\section{Proposed interpretive rules for a layered workplace battery}

The following design rules motivate the proposed battery. Their use in placement requires validation of the instruments, scoring, and decision thresholds in the target setting:

\begin{enumerate}[leftmargin=*]
    \item Use an objective test when claiming demonstrated foundation knowledge.
    \item Use self-report when confidence, perceived capability, trust, dependency, or reported practice is the intended construct.
    \item Add an ecologically grounded verification task when oversight of AI-generated claims matters.
    \item Keep trust and reliance separate. Trust can explain a reliance decision without proving that the decision was appropriate.
    \item Use a situated performance task for authority, state, recovery, independent review, and closure.
    \item Avoid one compensatory total across domains until empirical work demonstrates that such a total has a stable meaning.
\end{enumerate}

\end{document}